\documentclass{aa}

\usepackage{graphicx}
\usepackage{txfonts}
\newcommand{\onvire}[1]{}

\newcommand{\beq}{\begin{equation}}
\newcommand{\eeq}{\end{equation}}

\usepackage{color}
\usepackage{epsfig}
\usepackage{amssymb}
\usepackage{graphicx,natbib}
\usepackage{pdflscape}
\usepackage{longtable}
\usepackage{multirow}
\usepackage[figuresright]{rotating} 
\begin{document}

\title{The twofold debris disk around HD\,113766\,A\thanks{Based on {\it Herschel} observations, OBSIDs: 1342227026, 1342227027, 1342237934, and 1342237935. {\it Herschel} is an ESA space observatory with science instruments provided by European-led Principal Investigator consortia and with important participation from NASA. Based on \textsc{Visir} observations collected at the VLT
    (European Southern Observatory, Paranal, Chile) with program 089.C-0322(A)}}
  \subtitle{Warm and cold dust as seen with VLTI/\textsc{Midi} and Herschel/\textsc{Pacs}}

   \author{J. Olofsson
          \inst{1}
          \and 
          Th. Henning\inst{1}
          \and
          M. Nielbock\inst{1}
          \and 
          J.-C. Augereau\inst{2}
          \and 
          A. Juh\`asz\inst{3}
          \and
          I. Oliveira\inst{4}
          \and
          O. Absil\inst{5}
          \and 
          A. Tamanai\inst{6}
          }

   \offprints{olofsson@mpia.de}

   \institute{Max Planck Institut f\"ur Astronomie,
     K\"onigstuhl 17, 69117 Heidelberg, Germany \\
     \email{olofsson@mpia.de}
    \and
     UJF-Grenoble 1/CNRS-INSU, Institut de Plan\'etologie et d'Astrophysique de Grenoble (IPAG) UMR 5274, Grenoble, France
    \and
     Leiden Observatory, Leiden University, PO Box 9513, NL-2300 RA Leiden, The Netherlands
    \and
      Astronomy Department, University of Texas at Austin, 1 University Station C1400, Austin, TX 78712-0259, USA
    \and
      D\'epartement d'Astrophysique, G\'eophysique et Oc\'eanographie, Universit\'e de Li\`ege, 17 All\'ee du Six Ao\^ut, B-4000 Sart Tilman, Belgium
    \and 
      University Heidelberg, Kirchhoff-Institut f\"ur Physik, 69120 Heidelberg, Germany
   }

   \date{Received \today; accepted }

 
   \abstract
   {Warm debris disks are a sub-sample of the large population of debris disks, and display excess emission in the mid-infrared. Around solar-type stars, very few objects ($\sim$\,2\% of all debris disks) show emission features in mid-IR spectroscopic observations, that are attributed to small, warm silicate dust grains. The origin of this warm dust can possibly be explained either by a recent catastrophic collision between several bodies or by transport from an outer belt similar to the Kuiper belt in the Solar System.}
   {We present and analyse new far-IR Herschel/\textsc{Pacs} photometric observations, supplemented by new and archival ground-based data in the mid-IR (VLTI/\textsc{Midi} and VLT/\textsc{Visir}), for one of these rare systems: the 10--16\,Myr old debris disk around HD\,113766\,A. We improve an existing model to account for these new observations.}
   {We implemented the contribution of an outer planetesimal belt in the \textsc{Debra} code, and successfully used it to model the SED as well as complementary observations, notably \textsc{Midi} data. We better constrain the spatial distribution of the dust and its composition.}
   {We underline the limitations of SED modelling and the need for spatially resolved observations. We improve existing models and push further our understanding of the disk around HD\,113766\,A. We find that the system is best described by an inner disk located within the first AU, well constrained by the \textsc{Midi} data, and an outer disk located between 9--13\,AU. In the inner dust belt, our previous finding of Fe-rich crystalline olivine grains still holds. We do not observe time variability of the emission features over at least a 8 years time span, in a environment subjected to strong radiation pressure.}
   {The time stability of the emission features indicates that $\mu$m-sized dust grains are constantly replenished from the same reservoir, with a possible depletion of sub-\,$\mu$m-sized grains. We suggest that the emission features may arise from multi-composition aggregates. We discuss possible scenarios concerning the origin of the warm dust observed around HD\,113766\,A. The compactness of the innermost regions as probed by the \textsc{Midi} visibilities, as well as the dust composition, suggest that we are witnessing the outcomes of (at least) one collision between partially differentiated bodies, in an environment possibly rendered unstable by terrestrial planetary formation.}
   \keywords{Stars: individual: HD\,113766\,A – planetary systems: Zodiacal dust – circumstellar matter – Infrared: stars – Techniques: spectroscopic - high angular resolution}
\authorrunning{Olofsson et al.}
\titlerunning{The debris disk around HD\,113766\,A}

   \maketitle
%

\section{Introduction\label{sec:intro}}

Debris disks are byproducts of the planet formation process, once the massive, gas-rich circumstellar disk has dissipated. After the gas disks are lost, planetary systems including asteroid and comets may remain and evolve through gravitational interactions. The presence of dust, that originates from collissions or evaporation of these bodies, translates into a mid- to far-infrared (IR) excess, as demonstrated for the first time for Vega (\citealp{Aumann1984}), using the Infrared Astronomical Satellite. Departure from photospheric emission at these wavelengths is the consequence of thermal emission arising from dust grains heated by the stellar radiation. Several hundreds of stars are now known to harbor such excess emission (e.g., \citealp{Chen2006}). Our understanding of their nature has progressed during the past decades, both observationally and theoretically (see \citealp{Wyatt2008} and \citealp{Krivov2010} for recent reviews). In most cases, these systems compare very well with the Kuiper belt observed in our Solar System (e.g., \citealp{Carpenter2009}, \citealp{Liseau2010}, \citealp{Lohne2012}), where excess emission is detected in the far-IR. Several studies made use of high angular resolution instruments to detect hot exozodiacal dust around well-known debris disks (e.g., Fomalhaut and Vega, see \citealp{Absil2009}, \citealp{Defr`ere2011}, respectively). In such cases, the near-IR excess is relatively small ($\sim$\,1\%) and departure from photospheric emission only becomes significant at mid- or far-IR wavelengths. A small sample of the debris disk population is referred to as ``warm debris disks" (e.g., \citealp{Mo'or2009}) where emission in excess is  significant in the mid-IR, suggesting the presence of warm dust grains, close to the star. Within this class of warm debris disks, a few objects are of particular interest since they also display emission features in mid-IR spectroscopic observations. These emission features are associated to (sub-) $\mu$m-sized dust grains, that are warm enough to contribute significantly in the mid-IR (see \citealp{Henning2010} for a review on optical properties of silicate dust). 

These warm debris disks with emission features around Solar-type stars are rare ($\sim$\,2\% of all debris disks, \citealp{Bryden2006}, \citealp{Chen2006}). As discussed in \citet{Wyatt2008}, the abnormalous brightness of these debris disks cannot be solely explained by a steady state collisional evolution. If it was, the collisional lifetime of the parent bodies responsible for the dust production, and hence the infrared luminosity, would be too short compared to the age of the host star. Consequently, two scenarios are possible, either ({\it i}) a recent catastrophic collision between two planetesimals, or ({\it ii}) an outer belt feeding the inner regions, for instance via scattering of bodies by  one ore more planet (\citealp{Bonsor2012}) or in a similar scenario to the Late Heavy Bombardment phase, that happened in the Solar System (\citealp{Gomes2005}). The overall scarcity of the ``warm debris disk'' phenomenon would suggest a relatively low probability for such events to happen and produce enough dust to be detected. Thorough studies of such objects can therefore improve our understanding of the final steps of planetary formation, as well as provide key clues about the origin of zodiacal dust or Kuiper-belt dust in our solar system.

\object{HD\,113766} is one of these rare warm debris disks that show prominent emission features in the mid-IR. The system is a binary of two almost identical stars with spectral types F3/F5, with a projected separation of 1.3\,$''$ ($\sim$\,160\,AU at $d_{\star} = 123$\,pc, \citealp{2007ASSL..350.....V}), the dust being located around the primary HD\,113766\,A ($L_{\star} = 4.4 L_{\odot}$, \citealp{Lisse2008}). The Spitzer/\textsc{Irs} spectrum of the 10--16\,Myr old debris disk was first studied by \cite{Lisse2008}. In \citet{Olofsson2012} we presented a first model to reproduce the Spitzer/\textsc{Irs} spectrum of this source. In the following we present new Herschel/\textsc{Pacs} photometric observations at 70, 100 and 160\,$\mu$m, supplemented by archival VLTI/\textsc{Midi} interferometric measurements as well as (new and archival) VLT/\textsc{Visir} observations in the $N$- and $Q$-bands. In section\,\ref{sec:obs} we describe the observations and data processing, in Sect.\,\ref{sec:mod} we present our model and results. We discuss the implications of our findings in Sect.\,\ref{sec:discuss} before concluding.

\section{Observations and data processing\label{sec:obs}}

\subsection{Herschel/\textsc{Pacs} observations}

\begin{figure*}
\begin{center}
\includegraphics[width=1.95\columnwidth]{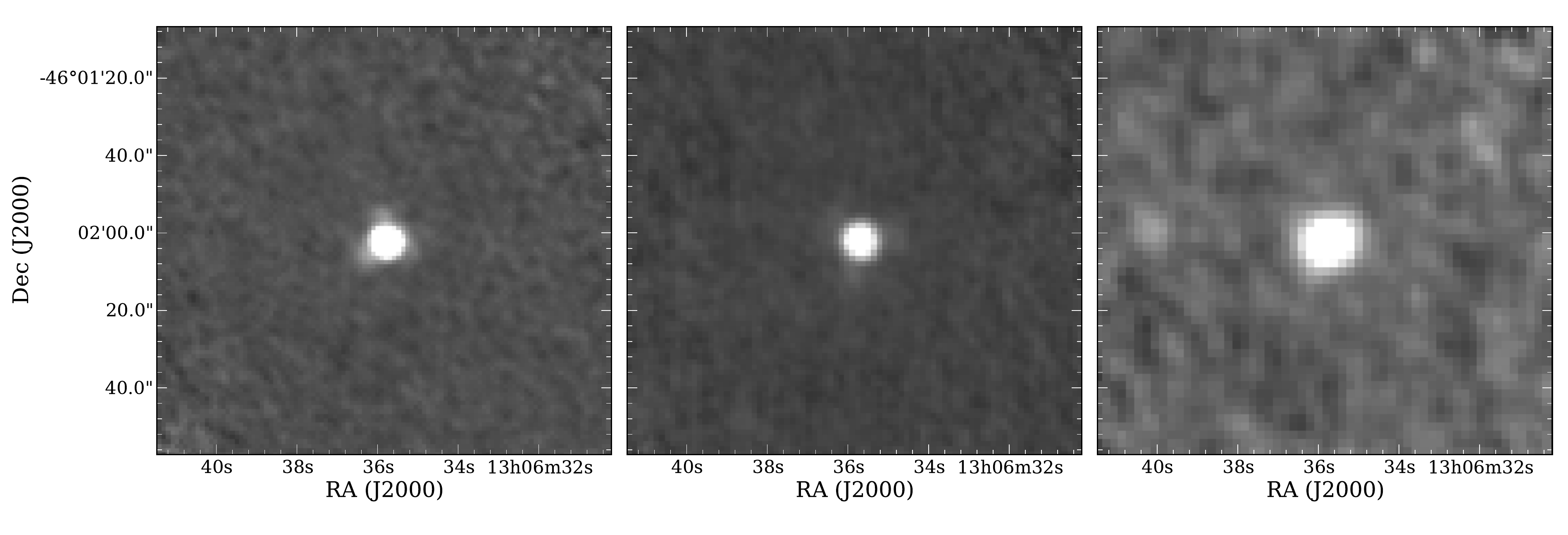}
\caption{Herschel/\textsc{Pacs} observations at 70, 100, and 160\,$\mu$m (left to right, respectively).\label{fig:data_pacs}}
\end{center}
\end{figure*}

We observed HD\,113766\,A with the \textsc{Pacs} photometer (\citealp{Poglitsch2010}) onboard Herschel (\citealp{Pilbratt2010}). The dataset consists of 4 \textsc{Pacs} mini-scan maps (program ``OT1\_jolofsso\_1''), two in the blue filter (70\,$\mu$m, OBSIDs 1342227026 and 1342227027), and two in the green filter (100\,$\mu$m, OBSIDs 1342237934 and 1342237935). The observing duration was $220 \times 2$ sec in the blue filter and $670 \times 2$ sec in the green filter. When observing with \textsc{Pacs} with the blue camera (blue or green filters, at 70 and 100\,$\mu$m, respectively), data is also simultaneously acquired with the red camera (160\,$\mu$m). We therefore combined the 4 mini-scan maps in the red filter to obtain the photometry at 160\,$\mu$m.

The data were processed using the Herschel interactive processing environment (HIPE, version 10.0.667), using standard scripts that include bad pixel flagging, detection and correction for glitches and HighPass filtering in order to remove the $1/f$ noise (see \citealp{Poglitsch2010} for more details). We discarded the first column of each detector sub-array in order to eliminate electronic crosstalk. Finally, we processed the data with different HighPass filter widths in order to test the robustness of the data reduction. The different values are tabulated in Table\,\ref{tab:flux} (column ``HP'').

The aperture photometry in all three filters was extracted within HIPE from the combined maps. The aperture radii are set at 10$''$ for the blue and green filters and 20$''$ for the red filter. The final flux values do not depend on the aperture sizes since an aperture correction is performed to compensate for flux loss outside of the aperture. This correction is based on a calibrated encircled energy fraction parametrisation contained in the calibration data that is included in HIPE. In order to estimate the noise level and thus the quality of the detection in all filters, we randomly placed apertures within a distance of 100, 75, and 50 pixels (for blue, green, and red filters, respectively) to the central pixel. Doing so, we avoided borders of maps that are more noisy due to a smaller detector coverage, and are therefore not representative of the noise close to the source.  We extracted the photometry for all of these apertures and then fitted a Gaussian profile to the flux distribution histogram to estimate the $\sigma$ uncertainty from the width of the Gaussian. Overall, these uncertainties are consistent with predictions from the \texttt{HerschelSpot}\footnote{http://herschel.esac.esa.int/Tools.shtml version 6.2.0} software. Figure\,\ref{fig:data_pacs} show the \textsc{Pacs} images at 70, 100, and 160\,$\mu$m (left to right, respectively), where the source is clearly detected at all three wavelengths. Table\,\ref{tab:flux} summarizes the final results of the data processing, with pixel sizes for the three filters, aperture corrected fluxes for the different HighPass filter widths, and estimated 1\,$\sigma$ uncertainties in units of mJy. The fluxes used in the analysis are the averages of those reported in Table\,\ref{tab:flux}. For all three filters, the photometric values are almost identical for the different HighPass filter widths, within the uncertainties. One should notice that the measured 70\,$\mu$m flux is consistent with the \textsc{Mips} observations reported by \citet[][$F_{70} = $390\,mJy $\pm$ 11\,mJy]{Chen2011}.

\begin{table}
\caption{Final products of Herschel/\textsc{Pacs} data processing. ``HP'' corresponds to the HighPass filter widths in units of number of readouts.\label{tab:flux}}
\begin{center}
\begin{tabular}{lccccc}
\hline  \hline
Filter & Wavelength & Pixel size & HP & Flux  \\
 & $[$$\mu$m$]$ & [$''$] & & [mJy] \\ \hline
  & &            & 15 & 386.5\,$\pm$\,7.2 \\
Blue  & 70 & 1.1 & 20 & 390.1\,$\pm$\,8.7 \\
  & &            & 35 & 391.4\,$\pm$\,9.7 \\
 \hline
 & &              & 20 & 223.4\,$\pm$\,5.6 \\
Green & 100 & 1.4 & 30 & 225.4\,$\pm$\,6.6 \\
 & &              & 45 & 226.7\,$\pm$\,7.5 \\
 \hline
   & &            & 35 & 94.3\,$\pm$\,11.1 \\
Red   & 160 & 2.1 & 45 & 99.3\,$\pm$\,10.1 \\
   & &            & 70 & 113.6\,$\pm$\,12.6 \\
\hline
\end{tabular}
\end{center}
\end{table}

\subsection{Complementary observations}

In addition to the far-IR observations, we included the Spitzer/\textsc{Irs} spectrum of HD\,113766\,A (between 5 and 35\,$\mu$m) in the analysis (see \citealp{Olofsson2012} for details on the data processing). We supplemented these data with VLT/\textsc{Visir} observations (our program and data published in \citealp{Smith2012}) and VLTI/\textsc{Midi} interferometric measurements (\citealp{Smith2012}). 

\begin{figure*}
\begin{center}
\includegraphics[width=.66\columnwidth]{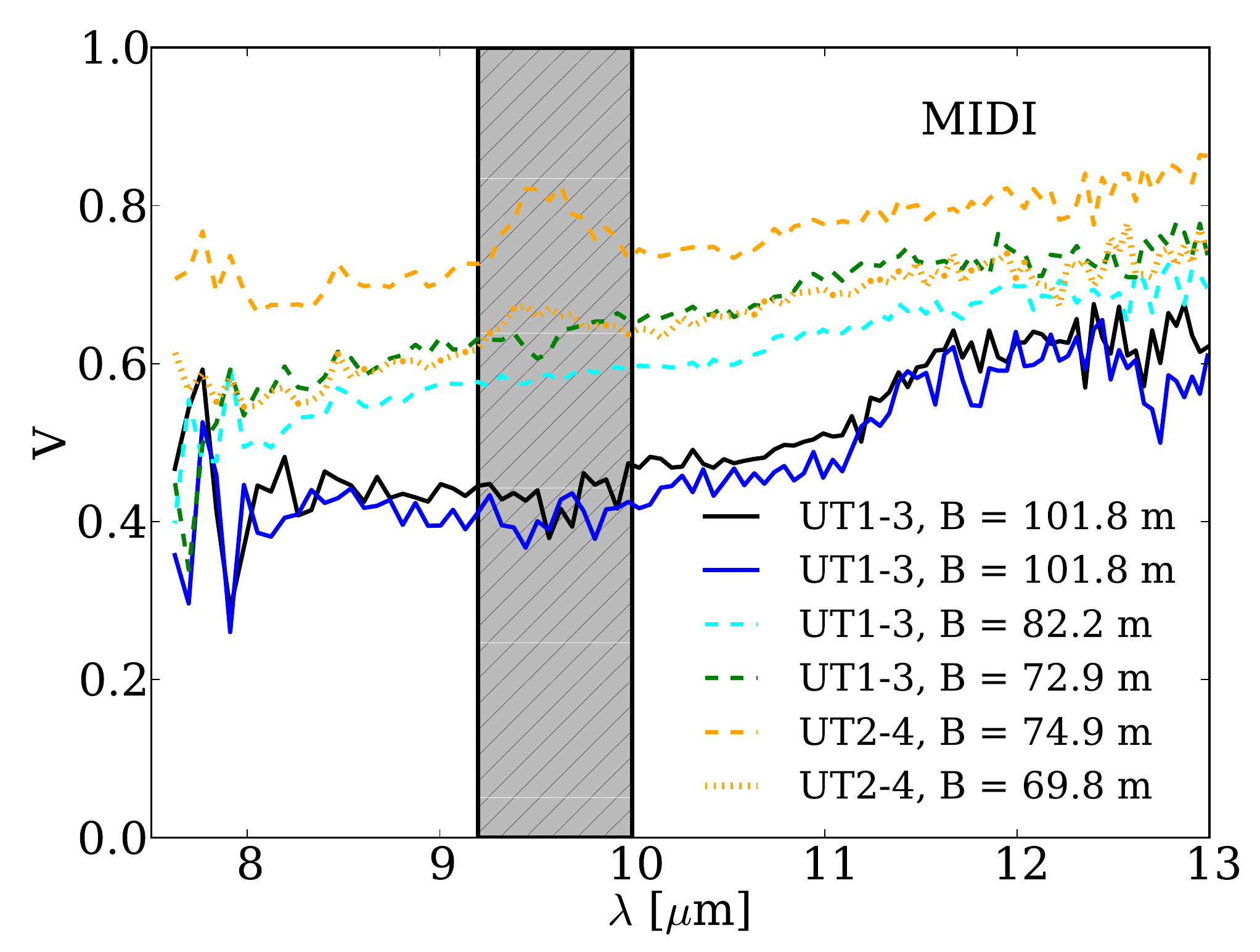}
\includegraphics[width=.66\columnwidth]{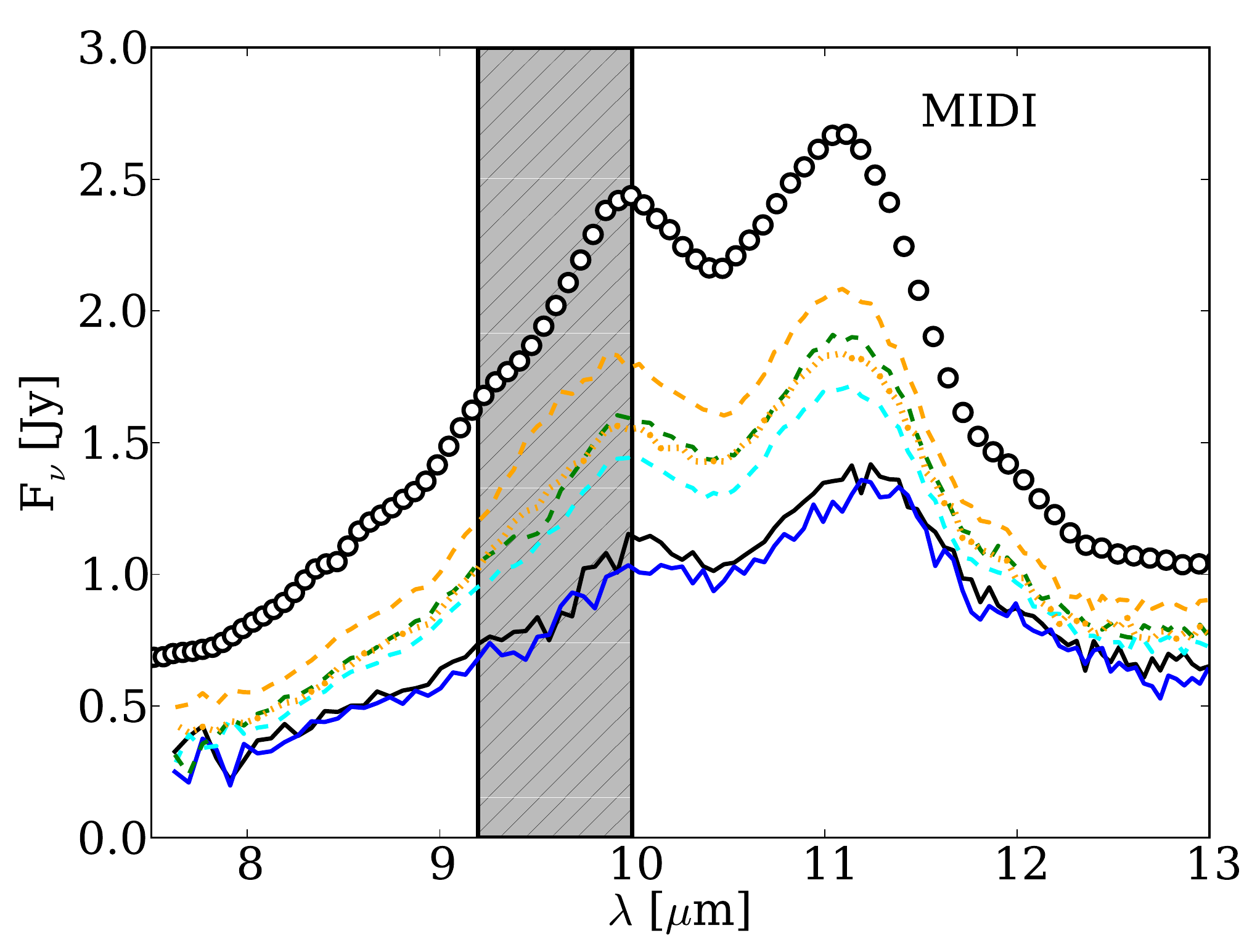}
\includegraphics[width=.66\columnwidth]{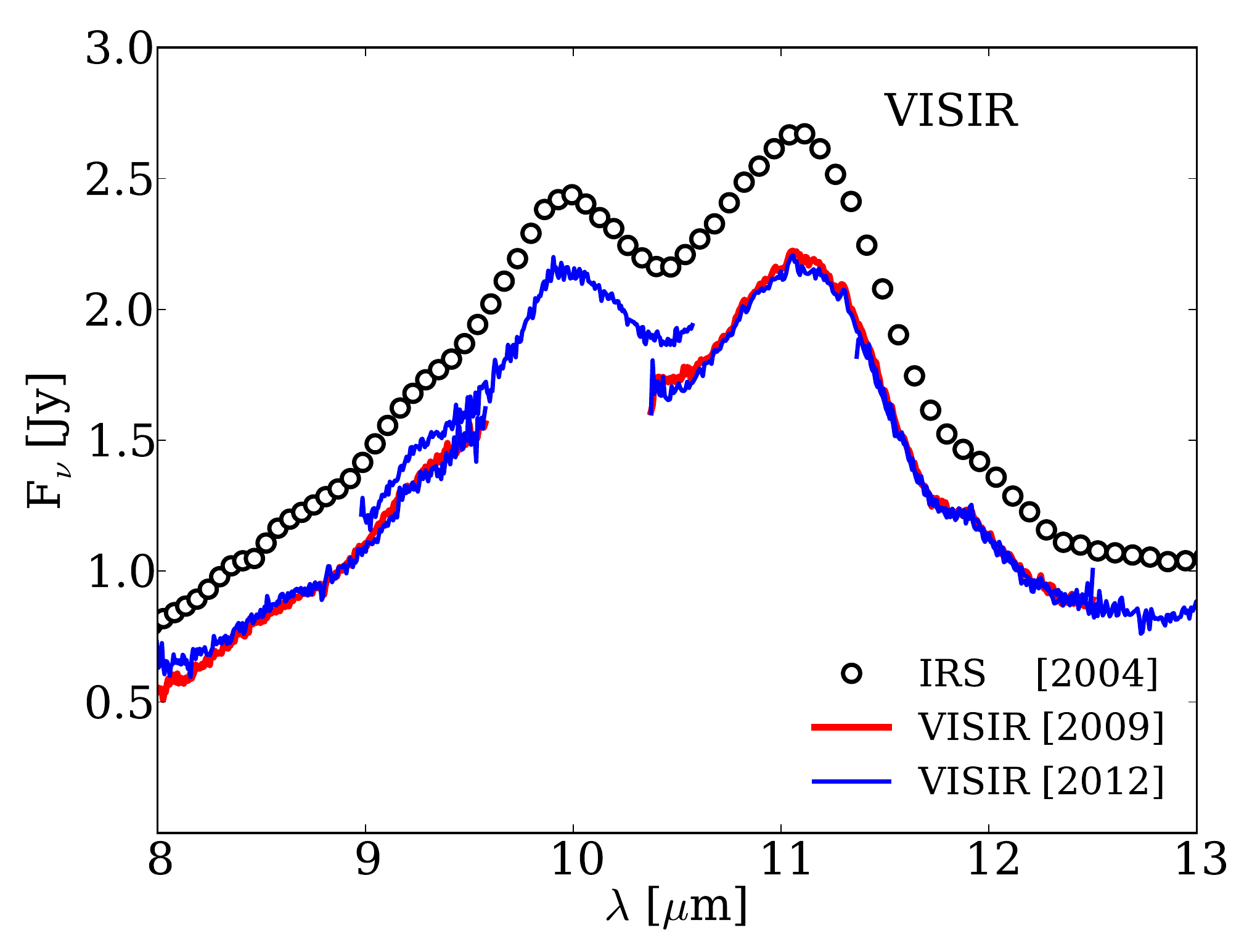}
\caption{VLTI/\textsc{Midi} visibilities and correlated fluxes for HD\,113766\,A (left and middle panels, respectively). The baselines used and their lengths are reported in the caption of the left panel. Right panel shows the \textsc{Visir} $N$-band spectrum at different epochs (2009 and 2012, in thick red and thin blue, respectively). The Spitzer/\textsc{Irs} spectrum is shown in open black circles in the middle and right panels.\label{fig:data}}
\end{center}
\end{figure*}

On the night of the 2012-Apr-21, HD\,113766\,A was observed with the VLT/\textsc{Visir} instrument (program 089.C-0322(A)) in the $N$-band. We defined four spectral windows, centered at 8.8\,$\mu$m ($8.0 \leq \lambda \leq 9.6 \,\mu$m), 9.8\,$\mu$m ($9.0 \leq \lambda \leq 10.6 \,\mu$m), 11.4\,$\mu$m ($10.4 \leq \lambda \leq 12.5 \,\mu$m), and 12.4\,$\mu$m ($11.4 \leq \lambda \leq 13.5 \,\mu$m), in low resolution mode ($R \sim 350$ at 10\,$\mu$m). We additionally retrieved the spectroscopic observations presented and analysed in \citet{Smith2012}. These observations were performed on the night of the 2009-May-11 in low resolution mode as well, for two spectral windows centered at 8.8 and 11.4\,$\mu$m (program 083.C-0775(E)). Both datasets were processed using standard scripts provided within the ESO Gasgano software\footnote{http://www.eso.org/sci/software/gasgano/ version 2.4.0}. Right panel of Figure\,\ref{fig:data} shows the \textsc{Irs} and \textsc{Visir} spectra obtained in 2004, 2009, and 2012 (open black circles, thick red, and thin blue lines, respectively). One should note that both \textsc{Visir} spectra, taken almost three years apart agree surprisingly well. However, the spectrum obtained in our program in the 9.8\,$\mu$m spectral window displays an offset of about 10\% higher in flux. Such ``stiching'' problems are not unusual, but the exact causes are not clear. We checked the atmospheric conditions during our observing program: the seeing and atmospheric pressures appear to be constant over the run. Besides the 9.8\,$\mu$m band offset, both \textsc{Visir} spectra show a deficit of flux of about 30\% compared to the \textsc{Irs} spectrum. \citet{Smith2012} already discussed in detail these differences (e.g., spatially extended emission) and concluded these differences to be within the absolute calibration uncertainites of \textsc{Visir} (see discussion in \citealp{Geers2007}).

The archival \textsc{Midi} data presented in \citet{Smith2012} were processed using scripts dedicated to observations of faint targets with \textsc{Midi} (see \citealp{Burtscher2012} for more details) interfaced with the \texttt{MIA+EWS} software\footnote{http://home.strw.leidenuniv.nl/$\sim$jaffe/ews/ version 2.0}. Table\,\ref{tab:midi} summarizes the observing log for the interferometric measurements. As in \cite{Smith2012}, we found that the observations performed on the night of the 2007-May-30 (program 079.C-0259(F)) had a low signal to noise ratio, and we did not include them in our analysis. The calibrated visibilities and correlated fluxes for HD\,113766\,A are displayed in the left and middle panels of Fig.\,\ref{fig:data}. The \textsc{Irs} spectrum is also shown for comparison to the correlated fluxes. Because of the challenge of calibrating accurately the \textsc{Visir} spectra, we used the \textsc{Irs} spectrum to compute visibilities from the correlated fluxes. The calibrated dataset compares very well with the data reduction of \cite{Smith2012}, except for one single run (083.C-0775(B), observed at 05:32:55). When calibrating these observations with their corresponding calibrator (HD\,112213, observed at 05:15:45), we found a correlated flux much higher than the \textsc{Irs} spectrum. We constructed the transfer function based on all the calibrators observed during that night. At the time of the science observations, we found a variation of about a factor 2 compared to averaged transfer function over the rest of the night. This suggests that the observing conditions were strongly varying. We reduced these science observations with a later observation of the same calibrator (06:34:00). One can see in Fig.\,\ref{fig:data} that both visibilities and correlated fluxes are large for this run (orange dashed line) and that there is a fairly significant difference with the consecutive observations on the same night, with the same baseline (06:09:53, orange dotted line). As explained in \citet{Burtscher2012} uncertainties are given by the quadratic sum of the photon noise error delivered by the \texttt{MIA+EWS} software and the standard deviation of the transfer function over the night of observations. Overall, the final uncertainties are dominated by the latter one (between 5 and 10\%). For the ``problematic'' observing run (083.C-0775(B), time 05:32:55), the standard deviation of the transfer function over the night is closer to 25\%.

We did not reduce the VLT/\textsc{Visir} images analyzed in \citet{Smith2012}. We instead used the full width at half maximum (FWHM) presented in their study. In none of the $N$- and $Q$-band images ($\lambda_{\mathrm{c}} = 11.85$ and 18.72\,$\mu$m with $\Delta \lambda =2.34$ and 0.88\,$\mu$m, respectively) presented by the authors, the debris disk was spatially resolved compared to a reference PSF. They fitted the azimuthally averaged surface brightness distributions with Gaussian profiles with FWHM of 0.322\,$''$ and 0.498\,$''$ in the $N$- and $Q$-bands, respectively. During the analysis, we will check if synthetic images convolved with such PSFs are consistent with the measurements of \citet{Smith2012}. After convolving the synthetic images with the reference PSF, we will compare the surface brightness distribution to check for possible resolved emission in both bands.

\begin{table*}
\caption{Log of the VLTI/\textsc{Midi} observations and averaged visibilities in the range 10.5\,$\mu$m $\pm\,0.5$\,$\mu$m, with 1\,$\sigma$ uncertainties (see text for details, Sect.\,\ref{sec:mod}). The $(u,v)$ values are in m.\label{tab:midi}}
\begin{center}
\begin{tabular}{lccccc}
\hline \hline
Program & Date & Time & Baseline & $(u,v)$ & $V$ \\
\hline
079.C-0259(G) & 2007-Apr-08 & 01:11:11 & UT1--UT3 & $(4.9,101.7)$  & 0.48$\pm$0.03 \\
079.C-0259(G) & 2007-Apr-09 & 00:33:40 & UT1--UT3 & $(-4.7,101.7)$ & 0.45$\pm$0.03 \\
083.C-0775(D) & 2009-May-08 & 05:15:29 & UT1--UT3 & $(65.3,49.9)$  & 0.61$\pm$0.03 \\
083.C-0775(D) & 2009-May-08 & 06:11:31 & UT1--UT3 & $(61.8,38.7$   & 0.68$\pm$0.04 \\
083.C-0775(B) & 2009-May-09 & 05:32:55 & UT2--UT4 & $(67.2,-33.0)$ & 0.75$\pm$0.10\\
083.C-0775(B) & 2009-May-09 & 06:09:53 & UT2--UT4 & $(57.0,-40.3)$ & 0.67$\pm$0.04 \\
\hline
\end{tabular}
\end{center}
\end{table*}

\section{Multi-technics, multi-wavelengths modelling\label{sec:mod}}

As in \citet{Olofsson2012}, a given debris belt is fully described by four free parameters plus a dust model: the inner and outer radii of the dust belt ($r_{\mathrm{in}}$ and $r_{\mathrm{out}}$), a volume density exponent $\alpha$ ($\Sigma (r) = \Sigma_0 (r / r_0)^{\alpha}$, $\alpha < 0$) and an exponent $p$ for the grain size distribution (d$n(s) \propto s^p$d$s$, $p < 0$). The $\Sigma_0$ reference value is found by adjusting the total mass of the dust $M_\mathrm{dust}$, which in turn is obtained by fitting the SED. The relative abundances for the different dust species are also free parameters in the modelling and are found by fitting the SED and the emission features detected in the \textsc{Irs} spectrum. For each set of $r_{\mathrm{in}}$, $r_{\mathrm{out}}$, $\alpha$ and $p$, the fitting of the abundances is performed via a Levenberg-Marquardt algorithm on a linear combination of the thermal contributions of all the dust species considered. The best-fit disk model is found via a Monte-Carlo Markov Chain (MCMC) exploration of the parameter space (\texttt{emcee} package, \citealp{Foreman-Mackey2012}). All the uncertainties quoted in the next sections represent the standard deviations of all the values tested during the fitting process.

In the following, we will first present a model of the debris disk assuming that the dust is distributed in a single belt around the central star (referred to as model ``1B'' for ``one belt''). We will test the robustness of this model, comparable to the one presented in \citet{Olofsson2012}, with respect to the new far-IR photometric observations and mid-IR interferometric measurements. The Herschel data provide new constraints on the cold counterpart of the debris disk, while the mid-IR \textsc{Midi} data are more sensitive to the warmer dust and provide direct information on the geometry of the dust distribution. As a matter of fact, the interferometric observations suggest the region emitting in the $N$-band is compact ($\lesssim 10$\,mas, or 1.2\,AU in diameter, according to \citealp{Smith2012}). On the other side, our previous best-fit model described in \citealp{Olofsson2012} was found to be fairly extended ($r_\mathrm{out} \sim 12$\,AU, with a density profile with an exponent of $\alpha = -1.8$). We can therefore already anticipate that fitting the SED and the \textsc{Midi} data simultaneously will be challenging.

In order to compare our model to the \textsc{Midi} and \textsc{Visir} observations we updated the \textsc{Debra} code to produce synthetic images at different wavelengths. To do so, at a given wavelength, we distribute the thermal surface brightness distribution ($F_{\nu}(r)$, between $r_{\mathrm{in}}$ and $r_{\mathrm{out}}$), assuming the disk is axisymmetric and infinitely flat. Possible inclinations effects are taken into account by projecting pixels of the image onto the sky plane. However, because we assumed the disk to be infinitely flat, we chose the inclination $i$ ($0^{\circ}$ being face-on) to be smaller than $60^{\circ}$. In the case of HD\,113766\,A, this choice is supported by the analysis of \citet{Smith2012} who could successfully model the \textsc{Midi} data with a circularly symmetric flux distribution (i.e., small $i$). As we are solely interested in the variation of the visibilities as a function of the baseline lengths, we averaged the \textsc{Midi} visibilities in the spectral range 10.5$\pm$0.5\,$\mu$m for each \textsc{Midi} baseline. We tested different values for $i$ and the position angle ($PA$) of the debris disk. We tested four different values for $i$ (0, 20, 40, and 60$^{\circ}$) and 30 values linearly distributed between 0 and 180$^{\circ}$ for the $PA$ ($0^{\circ}$ meaning the semi-major axis is oriented along the North-South direction).

\subsection{One belt (model ``1B'')\label{sec:1b}}

\begin{figure*}
\begin{center}
\includegraphics[width=2.1\columnwidth]{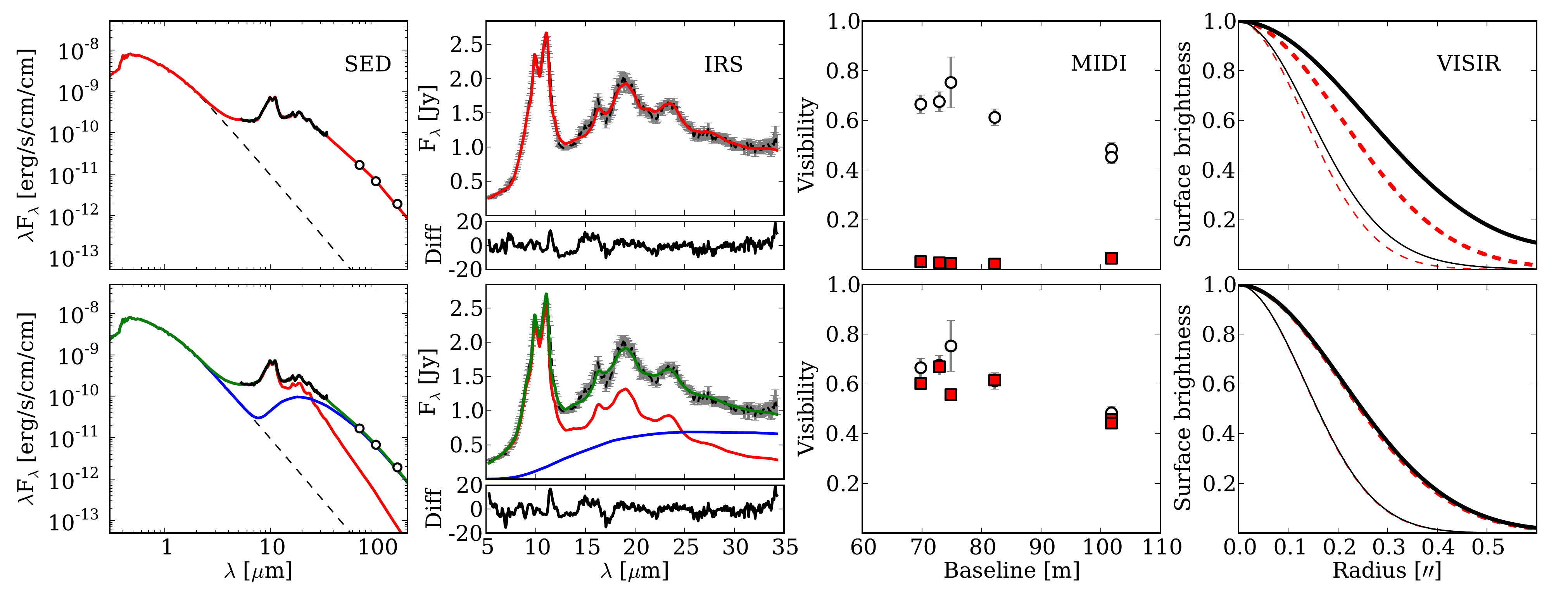}
\caption{{\it Upper and lower panels:} Results for the ``1B'' and ``2B'' models. {\it Left panel:} SED of HD\,113766\,A. The \textsc{Irs} spectrum is in black, \textsc{Pacs} data are shown as black open circles. For the ``1B'' model, the final best fit is in red. For the ``2B'' model, the final total fit is in green, the separate contributions of the inner and outer belts are in red and blue, respectively. {\it Middle left panel:} blow-up of the stellar subtracted \textsc{Irs} spectrum. Color coding is the same as for the left panel. Residuals ($100 \times [F_{\nu,\mathrm{obs}} - F_{\nu,\mathrm{mod}}]/F_{\nu,\mathrm{obs}}$) are displayed below the spectra. {\it Middle right panel:} Observed and modelled \textsc{Midi} visibilities as a function of the baseline (open black circles and red squares, respectively, with 1 $\sigma$ uncertainties). {\it Right panel:} modelled surface brightness at 10 and 20\,$\mu$m (thin and thick black lines, respectively) of synthetic images convoled with 2D Gaussian with FWHM of 0.322 and 0.498$''$. The 2D Gaussian with the aforementioned FWHM, assumed to be PSF representative of the images, are shown in thin and thick dashed red lines, for comparison.\label{fig:res}}
\end{center}
\end{figure*}

At first, we tried to model solely the SED, including the new \textsc{Pacs} photometric points, and the \textsc{Irs} spectrum. As mentioned earlier the \textsc{Midi} data suggest that the regions emitting in the $N$-band is compact ($r \leq 1.2$\,AU). Therefore, we first want to test how robust the assumption of a single dust belt is, with respect to the far-IR measurements, and underline at the same time the limitations of such approach. We will then compare the best-fit ``1B'' model to the interferometric data and discuss the subsequent implications on the spatial distribution of the dust. The dust properties (absorption efficiences and grains sizes) are the same as the ones described in \citet{Olofsson2012}. Grains sizes for the crystalline dust grains are $0.1 \leq s \leq 1$\,$\mu$m and $0.1$\,$\mu$m $ \leq s \leq 1$\,mm for the amorphous dust grains. The minimum and maximum values for the grain sizes are not free parameters in the modelling. This choice mostly originates from the experimental setup for the laboratory experiments that led to the extinction efficiencies of crystalline olivine grains (see \citealp{Olofsson2012} for more details).

Modelling only the SED with the new far-IR observations, we now find the dust to be located between ($0.4 \pm 0.2$)\,AU and ($51.9 \pm 25.2$)\,AU, with a volume density exponent of $\alpha = -1.9 \pm 0.3$, and an exponent for the grain size distribution $p = -3.0 \pm 0.1$. The total dust mass is $1.3 \times 10^{-2}\,M_{\oplus}$. The relative abundances for the different dust species are as follows: ($43.1 \pm 1.4$)\% of amorphous silicates (with a predominance of MgFeSi$_2$O$_6$ and MgFeSiO$_4$ grains, optical constant from \citealp{Dorschner1995}), ($17.3 \pm 0.7$)\% of Mg-rich crystalline olivine grains, ($21.0 \pm 0.6$)\% of Fe-rich crystalline olivine grains (laboratory measurements from \citealp{Tamanai2010} for the olivine data), ($11.6 \pm 0.5$)\% of enstatite (\citealp{Jaeger1998}), ($0.0 + 0.8$)\% of $\beta$-cristobalite (\citealp{Tamanai2010a}), and of ($7.0 \pm 0.6$)\% amorphous carbon dust grains (\citealp{Jager1998}, mostly in the form of cellulose pyrolized at 800$\degr$C, ``cel800''). Disk parameters for the best-fit model are reported in the first line of Table\,\ref{tab:disk}.

Fits to the SED, the stellar subtracted \textsc{Irs} spectrum, \textsc{Midi} data, and a comparison to the \textsc{Visir} images are displayed on the upper four panels of Figure\,\ref{fig:res} (left to right, respectively). For the SED, the \textsc{Irs} spectrum is shown in black, the \textsc{Pacs} data are shown as black open circles, and the best fit is represented as a solid red line. The color coding is the same for the blow up of the fit to the \textsc{Irs} spectrum (second column), and residuals are shown below the fit. Residuals are computed as $100 \times [F_{\nu,\mathrm{obs}} - F_{\nu,\mathrm{mod}}]/F_{\nu,\mathrm{obs}}$, where $F_{\nu,\mathrm{obs}}$ and $F_{\nu,\mathrm{mod}}$ are the observed and modelled fluxes, respectively. The third column panel shows the observed and modelled \textsc{Midi} visibilities for the best-fit to the SED (uncertainties are 1 $\sigma$). Right panel shows the comparison between the PSFs fitted to the \textsc{Visir} observations (\citealp{Smith2012}), in the $N$- and $Q$-bands and the azimuthally averaged profiles of synthetic images for the final best-fit model, convolved with PSFs with FWHMs of 0.322$''$ and 0.498$''$, respectively, as in \citet{Smith2012}.

While the best-fit model agrees well with the SED and the \textsc{Irs} spectrum, the observed \textsc{Midi} visibilities are severely under-predicted. As discussed above, this can be explained by the relative compactness of the emitting regions probed by the interferometric data. The $N$-band emission in our best-fit model is too extended and therefore the modelled visibilities drop very quickly. This is also true for the \textsc{Visir} observations. The modelled surface brightness profiles over-predict the observed 2D Gaussian profiles assumed to be representative of the PSFs, in the $N$-band and even more in the $Q$-band (the disk was not resolved in the \textsc{Visir} observations). 

As a final test, we tried to include the \textsc{Midi} observations in the fitting process (by summing the reduced $\chi^2_\mathrm{r}$ from the fits to the SED and \textsc{Midi} data), but could not achieve any better results. Both the \textsc{Midi} and the mid- to far-IR part of the SED are under-predicted. This can be explained as a tradeoff between the compactness of the dust belt and its contribution in the far-IR. Under the assumption of a single dust belt, the fits to the SED and interferometric data are diametrically opposed. From a modelling point of view, the main conclusion of this exercise is that SED modelling can be misleading: as shown in the upper left panel of Fig.\,\ref{fig:res}, we could obtain a decent fit to the SED, but this model is quickly proven wrong as soon as spatially resolved observations are available.

\subsection{Two belts (model ``2B'')\label{sec:2b}}

\begin{figure}
\begin{center}
\includegraphics[width=1\columnwidth]{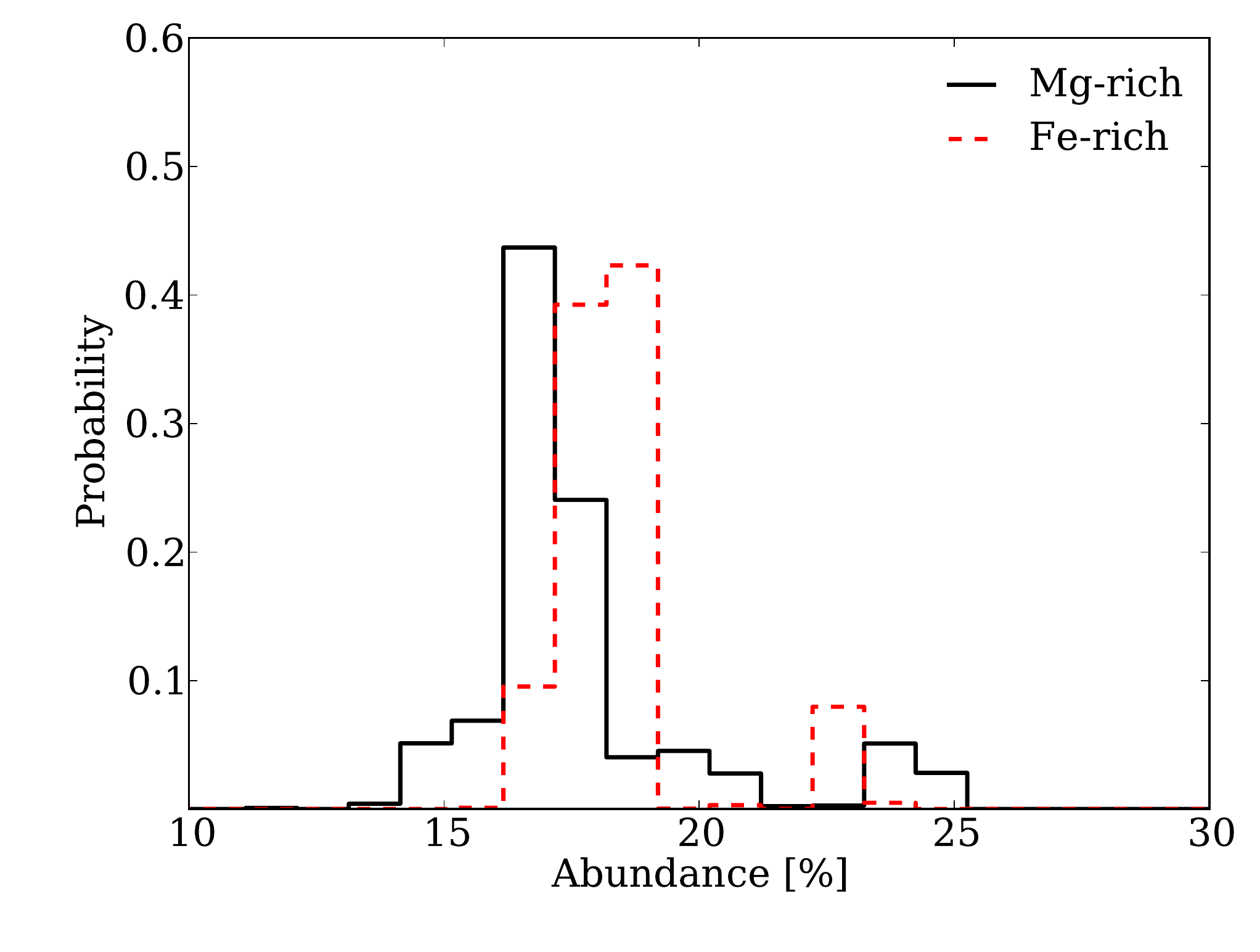}
\caption{Probability distributions from the MCMC run for model ``2B'', as a function of relative abundances of Mg- and Fe-rich olivine grains (solid black, and dashed red, respectively).\label{fig:Fe}}
\end{center}
\end{figure}

Several recent studies have shown that debris disks may contain several planetesimal belts (e.g., \citealp{Morales2011} or the disk around $\tau$ Ceti, \citealp{2007}; \citealp{Greaves2004}). We have shown in Sect.\,\ref{sec:1b} that the assumption of one single dust belt around HD\,113766\,A cannot account for all the available dataset. We can therefore start discussing the possibility of two, spatially separated dust belts around the central star: an inner dust belt, compact, close to the star that can reproduce the \textsc{Midi} data and the near- to mid-IR flux, and a second dust belt, farther away from the star that will contribute to the emission in the far-IR. To account for an additional belt, the modification to be applied to the \textsc{Debra} code is straightforward as we simply duplicated the computations done for one dust belt to the second one. A second radial grid, with different $r_\mathrm{in}$ and $r_\mathrm{out}$ values, with its own volume density ($\alpha$) is generated, as well as the corresponding dust temperatures as the function of the grain sizes and compositions. The dust properties in the outer belt can be defined separately from the ones in the inner belt (different dust species, $s_{\mathrm{min}}$, $s_{\mathrm{max}}$, and $p$). Finally, the fitting to the SED is performed over the sum of the contributions of the two dust belts (returning the dust mass for each belt).

\begin{table*}
\caption{Best-fit results for the disk parameters for both the ``1B'' and ``2B'' models\label{tab:disk}}
\begin{center}
\begin{tabular}{lccccccc}
\hline \hline
Model & Belt \# & $r_{\mathrm{in}}$ & $r_{\mathrm{out}}$ & $\alpha$ & $p$ & $M_{\mathrm{dust}}$ \\
      &         & $[$AU$]$          & $[$AU$]$           &          &     & $[M_{\oplus}]$  \\
\hline
One belt (``1B'')                  & 1 & $0.4 \pm 0.2$ & $51.9 \pm 25.2$ & $-1.9 \pm 0.3$ & $-3.0 \pm 0.1$ & $1.3 \times 10^{-2}$  \\
\hline
\multirow{2}{*}{Two belts (``2B'')} & 1 & $0.6 \pm 0.1$ & $0.9 \pm 0.3$ & $-1.7 \pm 0.3$ & $-3.5 \pm 0.1$ & $9.0 \times 10^{-5}$ \\
                        & 2 & $9.0 \pm 1.4$ & $13.1 \pm 3.4$ & $-1.2 \pm 0.3$ & $-3.8 \pm 0.2$ & $2.4 \times 10^{-3}$  \\
\hline
\end{tabular}
\end{center}
\end{table*}

As discussed in \citet{Olofsson2012}, the radiation pressure force is strong around HD\,113766\,A. We computed the unitless $\beta$ ratio between the radiation pressure and gravitational forces, for different grain sizes and dust compositions as described in \citet{Burns1979}. We find that grains with sizes smaller than ($4\pm0.5$)\,$\mu$m (depending on their compositions) are expected to be short-lived around the central star (see Sect.\,\ref{sec:time}). However, the detection of strong emission features indicates that warm (sub-)\,$\mu$m-sized grains must be located close to the star to contribute in the \textsc{Irs} spectrum. Therefore, the dust properties (absorption efficiencies and grain sizes) of the innermost dust belt in model ``2B'' (as in ``two belts'') are chosen to be similar to the ``1B'' model (Sect.\,\ref{sec:1b}), regardless of the radiation pressure issue. The relative abundances are still free parameters. For the outer dust belt, we did not include such small grains. The ratio $\beta$ between the radiation pressure and gravitational forces is distance-independent (both forces decreasing as $r^{-2}$, their ratio no longer depends on $r$), while the collisional timescale is longer at larger distances (Sect.\,\ref{sec:time}). We set the minimum grain size to $s_{\mathrm{min}} = 4$\,$\mu$m for all dust species in the outer belt, as there are no observational evidences, such as emission features at long wavelengths, that would require including smaller dust grains. Because we assumed in our model that crystalline dust grains have sizes smaller than 1\,$\mu$m, they are not accounted for in the outer belt (see Sect.\,\ref{sec:transient} for further discussion). This time, for each model of the MCMC run we fitted the SED and the \textsc{Midi} observations simultaneously.

For the inner dust belt, we obtained a best fit to the data with dust being located between ($0.6 \pm 0.1$)\,AU and ($0.9 \pm 0.2$)\,AU, a density profile in $\alpha = -1.7 \pm 0.3$, and a grain size distribution exponent of $-3.5 \pm 0.1$ (consistent with the prediction for a collisional cascade $p = -3.5$, \citealp{Dohnanyi1969}). The total dust mass is $9.0 \times 10^{-5}\,M_{\oplus}$. The relative abundances for the inner belt are the following: ($44.5 \pm 1.4$)\% of amorphous silicate grains (dominated by Mg$_2$SiO$_4$ grains, optical constant from \citealp{Jager2003}), ($17.0 \pm 0.8$)\% and ($18.8 \pm 0.8$)\% of Mg-rich and Fe-rich crystalline olivine grains, respectively, as well as ($7.7 \pm 1.2$)\% of enstatite, ($9.4 \pm0.4$)\% of $\beta$-cristobalite, and ($2.7 \pm 0.5$)\% of amorphous carbon grains (cellulose pyrolized at 800$\degr$\,C, ``cel800''). Figure\,\ref{fig:Fe} shows the marginalized probability distributions built from the MCMC run, for the relative abundances of Mg- and Fe-rich crystalline olivine grains (solid black and dashed red histograms, respectively). Even though more Fe-rich crystalline olivine grains are found compared to Mg-rich olivine, both distributions are overall close to each other. This result differs from the abundances previously found in \citet[][$\sim$\,4 times more Fe-rich olivine grains than Mg-rich crystalline grains]{Olofsson2012}. This difference can be explained as the outer planetesimal belt acts as a source of featureless, continuum emission. Consequently, this modifies the required contribution of the inner dust belt that is responsible for the emission features, and hence the relative abundances of the dust species. Nonetheless, it means that Fe-rich crystalline olivine dust grains are necessary to provide a good fit to the emission features detected in the mid-IR spectrum. 

For the outer belt, we found $r_{\mathrm{in}} =$ ($9.0 \pm 1.4$)\,AU and $r_{\mathrm{out}} = $ ($13.1 \pm 3.4$)\,AU, with $\alpha = -1.2 \pm 0.3$, and $p=-3.8 \pm 0.2$. The dust mass in the outer planetesimal belt is of about $2.4 \times 10^{-3}\,M_{\oplus}$. The dust content of the outer planetesimal belt consists entirely of amorphous carbon grains (68.7\% of ``cel600'' and 31.3\% of ``cel1000''). This can be explained by our choice of $s_\mathrm{min} =4$\,$\mu$m which damps most of the emission features, as well as the lack of significant emission features in the spectral range where the outer dust belt contributes most. Similar results were found by \cite{Lebreton2012} in the process of modelling the SED of HD\,181327. When considering only silicates and carbonaceous grains, carbon grains were preferentially used to fit the data. The authors pushed their models further by including the effects of porosity as well as the contribution of icy grains, reducing at the same time the relative fraction of carbon grains. For HD\,113766\,A, given the intertwined contributions of both the inner and outer belt, and the limited spectral coverage at wavelengths longer than 160\,$\mu$m, we did not perform similar investigations. One should also note that our result for the location of the outer belt is highly dependent on the minimum grain size $s_\mathrm{min}$ opted in our model. We find typical temperatures for the carbon grains between 100 and 200\,K. To have comparable temperatures, if $s_\mathrm{min} > 4$\,$\mu$m, grains will have to be located closer to the star, and vice-versa if $s_\mathrm{min} < 4$\,$\mu$m. Spatially resolved observations of the outer belt are mandatory to better constrain its location and the dust content. Even though the disk is not detected in the \textsc{Visir} images, they provide an upper limit of about 0.14$''$ (17\,AU at 123\,pc) for the maximal extent of the disk (\citealp{Smith2012}). Disk parameters for the best-fit model are reported in the last two lines of Table\,\ref{tab:disk}.

The lower four panels of Fig.\,\ref{fig:res} show the results for the ``2B'' model, with the same arrangement as described in Sect.\,\ref{sec:1b}. For the left and middle left panels (SED and \textsc{Irs} spectrum) the total fit is shown as a solid green line, while the two contributions of the inner and outer dust belts are represented as red and blue solid lines, respectively. Compared to the ``1B'' model, the best-fit model is much more compact and therefore both the \textsc{Midi} and \textsc{Visir} data are much better reproduced (except the ``problematic'' \textsc{Midi} run discussed in Sect.\,\ref{sec:obs}). As shown by the SED, the contribution of the outer belt in the $N$-band is almost negligible compared to the total flux. Therefore, the \textsc{Midi} observations are mostly sensitive to the inner belt, which is much more compact. We find its inclination to be of $40 \degr$ with a position angle of $120 \degr$, however, one should note that the inclination is sampled by steps of $20 \degr$ only. The overall compactness of the inner dust belt also explains why the modelled $N$-band \textsc{Visir} synthetic image agrees with the observations, as most of the emission at 10\,$\mu$m arises from the inner belt. As for the $Q$-band image, \citet{Smith2012} found an upper limit at 0.14$''$ (17\,AU at 123\,pc) for the maximum extent of the disk, consistent with our $r_\mathrm{out} = 13$\,AU.

Our best fit model indicates that the inner and outer belt has to be spatially separated. However, we further pushed our modeling effort and tried to fit the dataset with a single dust belt that shows a ``break'' in the density distribution (i.e., a single but {\it not} continuous belt as opposed to Sect.\ref{sec:1b}). The motivation being to mimic an inward radial drift of dust grains, followed by the formation of a dust ring at the sublimation radius, as described in \citet{Kobayashi2009} for instance. We therefore run a new set of models, for which we force the inner radius of the {\it outer} belt to be the same as the outer radius of the {\it inner} belt (seven free parameters instead of eight). The best fit results for the SED and the \textsc{Midi} data could not reach the quality of the best-fit ``2B'' model. We find that the disk extends from 0.6\,AU up to 16.5\,AU, with a break in the density at about 0.7\,AU. The fit to the \textsc{Irs} spectrum is acceptable except for the red wing of the 10\,$\mu$m emission feature where the modeled flux is larger than the observed flux. Because the disk extends farther out than $\sim$\,1\,AU, the interferometric measurements are still under-predicted ($V \sim 0.5-0.6$). To compare our results with the calculations presented in \citet{Kobayashi2009}, we compute the number density $n$ as a function of the distance $r$. For the best-fit model, the break at the intersection between the two adjacents dust belts has to be extremely large: the quantity $n \times r$ is more than five orders of magnitude larger in the innermost regions ($r < 0.7$) compared to the outermost regions of the disk ($r > 0.7$). If the warm ring was the consequence of radial drift and dust sublimation, \citet{Kobayashi2009} predicted an enhancement factor smaller than 10 for silicate dust grains around a F-type star. We can therefore confidently rule this scenario out to explain the origin of the transient dust around HD\,113766\,A.

As mentioned in Sect.\,\ref{sec:intro}, the secondary of the binary system is located at 1.3$''$ of HD\,113766\,A (160\,AU at 123\,pc). Without a fully characterized orbital solution for the binary system it is challenging to discuss in detail the influence of the secondary on the debris disk. However, it is unlikely that it can have a major effect on the outer disc located at $\sim$\,13\,AU from the primary. The only possibility would be if the companion star was on a very eccentric orbit with a periastron passing close to the outer belt, but in this case the secondary would spend most of its time in remote regions close to its apoastron, leaving much time for the circumprimary disc to ``recover'' from the perturbations at periastron passages (\citealp{Thebault2010,2012A&A...537A..65T}).

\section{Discussion\label{sec:discuss}}

Results presented in Sect.\,\ref{sec:mod} show that the available dataset for HD\,113766\,A can be best reproduced under the assumption of two spatially separated dust belts. Even though the SED alone can be reproduced under the assumption of a single dust belt, the \textsc{Midi} observations proved extremely useful to constrain the spatial extent of the region responsible for the emission in the $N$-band. If the inner dust belt extends farther out than $\sim$\,1\,AU, the modelled \textsc{Midi} visibilities are greatly under-predicted at all baselines. Our model is in good agreement with the results of \citet{Smith2012} who modelled the \textsc{Midi} data with either a symmetric Gaussian model with a FWHM of 10\,mas (1.2\,AU in diameter) or a ring model with a radius of 6\,mas ($\sim$\,0.7\,AU). Our modelling results, supported by the new \textsc{Pacs} observations enabled us to go one step further and demonstrate the presence of an additional planetesimal belt farther out in the system.

\subsection{Time variability\label{sec:time}}

Even though the absolute calibration of the \textsc{Visir} spectra is challenging, one can already see from the right panel of Fig.\ref{fig:data} that the shape of the emission features has not changed between 2004, 2009, and 2012. The relative strengths, the widths, and peak positions of the features seem to remain the same over a time span of at least 8 years. Such a phenomenon appears to be also valid for other debris disks with strong emission features (\citealp{Olofsson2012}, \citealp{Johnson2012}), while rapid dissipation of circumstellar material has only been observed for one single object (\citealp{Melis2012}). If (sub-)\,$\mu$m-sized crystalline dust grains are solely responsible for the emission features observed at 10 and 11\,$\mu$m, it would suggest that this population of grains is replenished from the very same reservoir (same composition), on the timescale of the evacuation by radiation pressure. To compute this timescale $t_\mathrm{rp}$, we followed the formalism to be presented in Lebreton et al. (in prep, private communication). The authors define $t_\mathrm{rp}$, which depends on the $\beta$ ratio, as the time needed for a dust grain to be transported from its initial distance $r$ to twice this distance $2 \times r$. As mentioned earlier, we computed the $\beta$ ratio for grains of different compositions as a function of their sizes. For HD\,113766\,A, this ratio peaks at $\beta = 3$ for grains with sizes of about 0.5\,$\mu$m. For these dust grains, assuming $r=0.75$\,AU ([$r_{\mathrm{out}} - r_{\mathrm{in}}]/2$), we find $t_\mathrm{rp} \sim 1$\,month. On the other hand, small dust grains are continuously produced by collisions. To estimate the collisional lifetime of the smallest grains in the inner dust belt, we  used the formula $t_\mathrm{c} = (\Omega \tau)^{-1}$, where $\Omega$ is the angular velocity at a distance $r$ and $\tau$ is the geometrical vertical optical depth (\citealp{Thebault2007}). Even though the exact collisional timescales may depart from this value, it provides an order of magnitude estimate. At a distance $r=0.75$\,AU, we find a collisional lifetime - source of dust replenishment - of about 2 months, twice the timescale of evacuation by radiation pressure $t_\mathrm{rp}$. For the outer belt, assuming $r = 10$\,AU, we find $t_\mathrm{rp} \sim 5$\,years (for grains with $\beta = 3$) and $t_\mathrm{c} = 300$\,years, which supports our choice of $s_\mathrm{min} = 4$\,$\mu$m in the outer belt. One should note that the timescale for the Poynting-Robertson drag, is much longer compared to the aforementioned timescales and can be neglected in the case of HD\,113766\,A (\citealp{Wyatt2005}).

The sub-\,$\mu$m-sized grains responsible for the observed emission features should be evacuated rapidly from the inner belt. One may therefore consider dust aggregates of different compositions (e.g., \citealp{Lindsay2010}) as a possible explanation. Computations of absorption coefficients, making use of the discrete-dipole approximation method (\citealp{Draine1988}) have shown that small crystalline monomeres may be incorporated into larger dust aggregates but still show prominent emission features associated with crystalline dust grains (e.g., \citealp{Wooden2012}). These larger aggregates would then be less subjected to radiation pressure and stay gravitationally bound to the central star on longer timescales. But even these aggregates will eventually collide, fragment, and be blown out of the system. This suggests that replenishment from a reservoir with the same bulk composition has to occur. Another way to explain the overall stability of the emission features would be to introduce a deficit in the grain size distribution for grains that are affected by radiation pressure (e.g., \citealp{Johnson2012}). However, this modelling is out of the scope of this paper given the total number of free parameters and the lack of spatially resolved observations to better constrain the location of the dust.

\subsection{The optical depth and opening angle}

Natively, the model space in the \textsc{Debra} code is 1D: there is only a radial dependence on the density and temperatures of the dust grains. However, an infinitely flat disk would quickly become radially optically thick. Therefore, the vertical dimension is described via an opening angle $\theta$ ($0\degr$ being an infinitely flat disk). At constant volume density, increasing $\theta$ decreases the local density in all of the radial cells, and hence the optical depth becomes smaller as well. For a given disk model, the \textsc{Debra} code returns a threshold value $\theta_\mathrm{lim}$, which is the smallest opening angle for which the debris disk is still in the optically thin regime. One should note that for small opening angles, $\theta$ is comparable to twice the aspect ratio of the disk $2 H/r$ (same formalism as in \citealp{Thebault2009}). For the best-fit ``2B'' model, we found a $\theta_\mathrm{lim,\,inner}$ of about 3.8$\degr$ for the innermost belt, a value consistent with the one found by \citet[][$4\degr$]{Lisse2008}. The opening angle for the outer belt is smaller ($\theta_\mathrm{lim,\,outer} \sim 0.9\,\degr$). The lower limit for the opening angle $\theta_\mathrm{lim,\,inner}$ is in good agreement with the work presented in \citet{Thebault2009}, who found a minimum ``natural'' observed aspect ratio of $\sim 2.3\,\degr$ (defined as $\theta /2$). 

\subsection{The origin of the transient dust\label{sec:transient}}

In \citet{Olofsson2012} we hypothesized that the transient dust grains observed around HD\,113766\,A may be the outcome of a disruptive collision of partially differentiated bodies. This conclusion was based on specific mineralogic markers: the composition of crystalline olivine grains containing a Fe/[Fe + Mg] fraction of about 20\%. Based on the new observations available, we refined our previous best-fit model by including an additional planetesimal belt farther away from the star. How does this new model fit in this picture ? One scenario that has been suggested in the literature to explain the excess emission observed around warm debris disks is that an outer belt is ``feeding'' the innermost regions. Comet-like bodies from the outer regions may be scattered inwards, for instance by a (chain of) planet(s) as described in \citet{Bonsor2012}. These bodies will then collide with each other or sublimate and release small dust grains in the immediate vicinity of the star. With our finding of an outer dust belt, one may wonder if such an event is taking place around HD\,113766\,A. 

The $N$-band \textsc{Midi} observations ruled out any significant emission to be outside of $\sim$\,1\,AU, suggesting that the phenomenon responsible for the transient dust has to be ``local'', as opposed to extended. For icy comets scattered inwards from the outer belt, sublimation may release a population of dusty grains. The majority of these grains would be released inside of 3.3\,AU, based on heating of large, km-sized comets by a stellar radiation field from a $4.4L_\odot$ star (private communication with Amy Bonsor and Ulysse Marboeuf, Bonsor et al. in prep). The fate of such grains released from a comet is unclear- whilst they could supply the observed excess emission, release of dust grains from an eccentric comet does not by itself explain the compact nature of the inner dust belt. If the observed warm material originated from the outer regions, the \textsc{Midi} data could only be explained by the capture of a massive comet that is out-gasing and evaporating close to the star (a local production of dust, \citealp{Beichman2005} concerning HD\,69830).

Our best-fit model can successfully reproduce the emission features detected in the \textsc{Irs} spectrum without including crystalline dust grains in the outer belt. This choice, initially motivated by the strength of the radiation pressure can also be justified by the lack of emission features associated with crystalline grains beyond 30\,$\mu$m. If the outer belt would contain small crystalline dust grains, they would be cold enough to be responsible for emission features at long wavelengths (especially the 33\,$\mu$m feature arising from crystalline olivine grains). The thermal emission of the outer belt peaks at about 25--30\,$\mu$m, and the fact that the 33\,$\mu$m feature is not detected suggests that such $\mu$m-sized crystalline grains are not present, at a detectable level, in the outer belt. The effect of crystalline monomeres inclusions in aggregates has been studied for the 10\,$\mu$m emission feature (e.g., \citealp{Min2008}), but not for other emission features at longer wavelengths. Further investigations have to be performed to estimate how much crystalline grains can potentially be included in aggregates that would still not show any detectable emission feature around 30\,$\mu$m. Overall, this suggests that the crystallinity fraction has to be smaller in the outer belt compared to the inner belt, pointing towards a different composition between the two dust belts. Would the mineralogical differences between the two belts be confirmed, this would argue against the capture of a massive comet originating from the outer regions.

Our spectral decomposition results suggest that the crystalline olivine grains in the inner belt contains more Fe than what is expected for comet-like bodies (\citealp{Wooden2008}; \citealp{Zolensky2008}; \citealp{Hanner2010}), cold crystalline olivine grains observed in other debris disks ($\beta$\,Pic, \citealp{2012}), or even younger proto-planetary disks (e.g., \citealp{Olofsson2010}; \citealp{Sturm2010}; Sturm et al. submitted, and \citealp{Juh'asz2010} who found an upper limit of 10\% for the Fe fraction). As previously discussed in \citet{Olofsson2012}, crystallization via thermal annealing or gas-phase condensation are unlikely to produce Fe-bearing crystalline olivine grains in proto-planetary disks (\citealp{Nuth2006}; \citealp{Gail2004}). Observing a significant fraction of Fe-rich crystals (more than half the total abundance of crystalline olivine grains) indicates that the dust composition has been altered. Internal metamorphism of the olivine content, as discussed in \citet{Nakamura2011} concerning the asteroid Itokawa, can provide an explanation for our finding. Consequently, with present-day observations and laboratory measurements of optical properties, we reiterate our previous conclusion that we may be witnessing the aftermath of a disruptive collision within the first AU from the central star. The inferred composition suggests that the dust must have been processed in a way which is not consistent with our current understanding and knowledge about the mineralogy of cometary bodies and proto-planetary disks. A collision of differentiated bodies as the origin of the transient dust seem to provide a better explanation of our results. If so, this conclusion also suggests the parent bodies do not originate from the outer belt, supporting the scenario of a ``local'' event. As discussed in \citet{Trieloff2006}, thermal metamorphism by internal heating is expected to be more efficient inside the first planetesimals that are formed, as they accrete more short-lived radionuclides ($^{26}$Al), while planetesimals that accreted material later on remain undifferentiated. As the timescale for planetesimal formation depends on the distance to the star (and the surface density, \citealp{Kenyon2004}, \citealp{Wyatt2008}) planetesimals in the outer belt are formed later on and thus may remain undifferentiated.

In a recent study, \cite{Jackson2012} investigated the outcome of a massive collision in the framework of the formation of the Moon. They found that such an impact would release enough dust for the corresponding excess emission to be detectable in the mid-IR up to 25\,Myr after the collision happened. Their numerical simulations suggest that the orbits of the debris released upon the impact would circularize in about 10\,kyr. Naturally, to take place, such event require the presence of large planetesimals, implying that km-sized bodies must have formed around HD\,113766\,A. As detailled in the review by \citet[][and references therein]{Wyatt2008}, planet formation models suggest that such bodies had sufficient time to be formed in the 10--16\,Myr old system. In turn, the chaotic phase of planetesimal growth in the innermost regions scatters the orbits of km-sized bodies and increases the probability of a collision (e.g., \citealp{Kenyon2006}).

\section{Conclusion}

In this study, we presented a new model for the debris disk around HD\,113766\,A. This model accounts for new far-IR photometric observations as well as archival mid-IR interferometric measurements. We underlined the limitations of SED modelling and the importance of spatially resolved observations. Our approach of modelling simultaneously the SED from near-IR to far-IR wavelengths and interferometric measurements has enabled us to have a deeper understanding of the morphology of the debris disk around HD\,113766\,A. We found that the available dataset can be best explained by two spatially separated dust belts. A first inner dust belt is located within 1\,AU from the star, a distance well constrained by the \textsc{Midi} observations. This inner belt contains warm sub-$\mu$m-sized dust grains. These small grains are responsible for the emission features observed via mid-IR spectroscopy. An outer planetesimal belt, located between 9--13\,AU (distances much less constrained by the observations) account for the far-IR emission observed with \textsc{Pacs}. The new observations support our previous conclusion (\citealp{Olofsson2012}), where we suggested that the observed transient dust is the result of a disruptive collision. This collision could be the consequence of on-going terrestrial planet formation around the young debris disk (10--16\,Myr).

\begin{acknowledgements}
The authors thank the referee, Philippe Th\'ebault, for positive and constructive comments that improved the paper. JCA thanks the French National Research Agency (ANR) for financial support through contract ANR-2010 BLAN-0505-01 (EXOZODI)
\end{acknowledgements}

\bibliography{biblio}

\end{document}